\documentclass{aa} 
\usepackage{txfonts}
\usepackage{graphicx} 
\usepackage{natbib} 
\bibpunct{(}{)}{;}{a}{}{,}

\begin{document} 

\title{Carbon monoxide in the solar atmosphere}
\subtitle{I. Numerical method and two-dimensional models}

\author{S.~Wedemeyer-B\"ohm \inst{1}
   \and I.~Kamp \inst{2}
      \and J.~Bruls \inst{1} 
      \and B.~Freytag \inst{3}
   }
 
\offprints{wedemeyer@kis.uni-freiburg.de} 
 
\institute{Kiepenheuer-Institut f\"{u}r Sonnenphysik, Sch\"{o}neckstra\ss e~6, 
     79104~Freiburg, Germany      
     \and Space Telescope Science Institute, 3700 San Martin Drive, Baltimore MD 21218, USA 
     \and Department for Astronomy and Space Physics, Uppsala University,  
     Box~515, 75120~Uppsala, Sweden 
   }
   \date{Received 15 December 2004 / Accepted 15 March 2005} 
\abstract{
The radiation hydrodynamic code \mbox{CO5BOLD} has been supplemented with the 
time-dependent treatment of chemical reaction networks. 
Advection of particle densities due to the hydrodynamic flow field is also 
included. 
The radiative transfer is treated frequency-independently, i.e. grey, so far. 
The upgraded code has been applied to two-dimensional simulations 
of carbon monoxide (CO) in the non-magnetic solar photosphere and low chromosphere. 
For this purpose a reaction network has been constructed, taking into account 
the reactions which are most important for the formation and dissociation of 
CO under the physical conditions of the solar atmosphere.
The network has been strongly reduced to 27 reactions, involving the chemical species 
H, H$_2$, C, O, CO, CH, OH, and a representative metal.
The resulting CO number density is highest in the  cool regions of the 
reversed granulation pattern at mid-photospheric heights and decreases strongly 
above. There, the CO abundance stays close to a value of  
8.3 on the usual logarithmic abundance scale with \mbox{[H]$=$12} 
but is reduced in hot shock waves which are a ubiquitous phenomenon of 
the model atmosphere. 
For comparison, the corresponding equilibrium densities have been calculated, 
based on the reaction network but also under assumption of instantaneous 
chemical equilibrium by applying the Rybicki \& Hummer (RH) code by \citet{uitenbroek01}. 
Owing to the short chemical timescales, the assumption holds for a large fraction 
of the atmosphere, in particular the photosphere. 
In contrast, the CO number density deviates strongly from the corresponding 
equilibrium value in the vicinity of chromospheric shock waves.  
Simulations with altered reaction network clearly show that the 
formation channel via hydroxide (OH) is the most important one under the 
conditions of the solar atmosphere. 

} 
\maketitle 
%
\section{Introduction} 
\label{sec:intro} 

Since \citet{noyes72b} deduced  very low brightness temperatures in the 
(upper) photosphere from observations in  the core of the carbon monoxide (CO) 
\mbox{3-2\ R14}  line,  
the thermal structure of this layer has been subject of an 
ongoing controversial debate.  
In many subsequent observations the low temperatures have not only been 
confirmed, e.g. by \citet{ayres81b}, but also found to be 
as low as 4000~K and even down to 
only 3700~K \citep[see Figs.~4-5 in][]{ayres81b}. 
These small values fall below the values implied by commonly used 
semi-empirical models like VAL-C 
\citep{val81} that are based on other diagnostics, e.g., in the UV range.   
Moreover, the CO observations did not show a prominent temperature inversion which is 
an important feature of VAL-C-like models.  
Ayres\ \&\ Testerman explained this 'diagnostic dilemma' with the 
thermal bifurcation of the outer solar layers. 
Following that idea, the hot plasma would be confined in discrete structures  
embedded in cool gas that can account for the observed carbon monoxide features. 
The inhomogeneities would be preferably of small spatial scale and large 
thermal contrast. 

A number of investigations imply that the bulk of CO is located in 
the low chromosphere and below.
For instance, Ayres\ \&\ Testerman state that the CO concentration 
peaks at an optical depth of $\tau_{500} \la 10^{-2}$ for a reference wavelength of 
$500$~nm. 
That is in line with the results by \citet{solanki94} who observed CO in emission 
at the solar limb. 
There, it remains at a constant level up to \mbox{$0$\,\farcs$4$} ($\approx 300$~km) above the 
(continuum) limb before it decreases rapidly in the layers above (see their Fig.~2). 
\citet{uitenbroek94} carried out similar observations,  
but with an improved technique, and found CO in 
emission, extending   \mbox{$0$\,\farcs$5$}  ($\approx 360$~km) beyond the continuum limb. 
\citet{ayres96} derive, based on the $\Delta V = 1$ bands, an off-limb 
extension of  \mbox{$0$\,\farcs$6$} ($\approx 440$~km) for the strongest lines, indicating the 
presence of cool gas up to $350$~km above the classical temperature 
minimum.


Given the high temperatures implied by other diagnostics, the above-mentioned CO 
observations inevitably demand for a spatially inhomogeneous structure of the 
photosphere and low chromosphere. 
In this sense, \citet{solanki94}, based on observed horizontal velocities, suggest 
that the bulk of CO might be located above granule interiors.  
Direct evidence for this assumption was provided by \citet[][ see also Uitenbroek et al., 1994]{uitenbroek00a}, 
who took spectroheliograms in the cores of CO lines.
They discovered not only a pattern connected to the magnetic network, 
but also bright rings with dark centres in the quiet Sun, resembling a 
reversed granular pattern. 
Moreover, they conclude that such spatial variations are largely of dynamic 
nature, and that dynamics thus play an important role in the formation of the 
dark CO line cores. 
In fact, oscillatory intensity changes \citep[see, e.g.,][]{ayres04} have already 
been detected by \citet{noyes72b}.


A number of static models have been constructed in order to explain the 
observations, ranging from pure 1D atmospheres like the COmosphere 
\citep{wiedemann94} to two-component models \citep[e.g.][]{ayres86, ayres96} 
and even more complex spatial configurations \citep[e.g.][]{ayres91, ayres02}.
But although indicating the right direction, the obvious dynamic nature of CO 
features cannot be reproduced by means of a static approach. 
Spatial \textit{and} temporal variations have to be taken into account 
for a detailed model. 

Instead of inverting observed intensities to a temperature stratification with 
one or more components, 
\citet{uitenbroek00a} started from a snapshot of the 3D  hydrodynamical model 
by \citet{stein89} and calculated CO number densities under the assumption 
of instantaneous chemical equilibrium (ICE). 
The highest concentration was found in the middle photosphere 
($\sim 100 - 300$~km above optical depth unity) above granule interiors, while  
the concentrations above intergranular lanes (at the same height) are  $3-4$ 
times smaller. 
The CO distribution is strongly connected to the cooling due to 
strong adiabatic expansion and cooling above the granule centres. 
The calculations were performed under the assumption of instantaneous
chemical equilibrium. 
Neither the radiative cooling action by CO itself nor 
advection due to the hydrodynamic flow were taken into account. 
This pioneering work showed the importance of solar granulation 
for CO line formation. 

\citet{asensio03} proceed to non-equilibrium CO chemistry calculations 
based on the time-dependent hydrodynamic simulations by \citet{carlsson97a}.
Solving the chemistry equations for a chemical reaction network, including 
the most relevant species, resulted in CO number density as function of height 
and time. 
From this Asensio Ramos et al. conclude that the radiation in CO lines close 
to the solar limb originates from heights not greater than 700~km. 
Moreover, they proved that ICE is a valid assumption for the lower 
layers but overestimates the CO number densities above the low chromosphere. 
The calculations were one-dimensional only and thus did not 
allow for an analysis of the horizontal distribution. 

This paper is the first part of a series. Here we present time-dependent 
CO chemistry as part of multi-dimensional radiation hydrodynamics simulations, 
starting with the 2D case. 
The major advance with respect to earlier investigations is the time-dependent 
treatment of a chemical network in combination with advection of particle densities 
with the hydrodynamic flow in two or 
three spatial dimensions. The upgraded code \textsf{\mbox{CO$^5$BOLD}} \citep{cobold} 
will serve as a base for further 
improvements as, e.g., the back reaction of CO as a cooling agent. 

After the description of the methods, the chemical input data, and the numerical 
model in Sects.~\ref{sec:method}, \ref{sec:network}, and \ref{sec:model}, respectively, 
we present the results of two-dimensional simulations in Sect.~\ref{sec:results} which 
are discussed in Sect.~\ref{sec:discus}. 
Finally, conclusions are drawn in  Sect.~\ref{sec:conclusion}.

\section{Numerical method} 
\label{sec:method} 

\subsection{Formulation of the problem}
\label{sec:problem}

The number density $n_i$ of a chemical species within a fixed volume can change
in time ($t$) due to  advection as expressed by the continuity equation: 
\begin{equation}
\frac{\partial{n_i}}{\partial{t}}\ +\ \nabla\cdot(n_i \vec{v}) = 0
\enspace,
\end{equation}
where $\vec{v}$ denotes the velocity of the hydrodynamic flow. 
An additional source term must be taken into account, if the  number density also 
changes due to chemical reactions. Such changes 
can be realised by a large variety 
of reactions. 
In the present application we restrict ourselves to two- and three-body 
reactions. The new term can then be written as 
\begin{eqnarray}
\label{eq:dndt}
\left(\frac{\partial{n_i}}{\partial{t}}\right)_\mathrm{chem} 
 &=& - n_i   \sum_j k_{2,ij}\ n_j \\
 & & + \quad \sum_j \sum_l k_{2,jl}\ n_j n_l                \nonumber\\
 & & - n_i   \sum_j \sum_l k_{3,ijl}\ n_j\ n_l               \nonumber\\
 & & + \quad \sum_j \sum_l \sum_m k_{3,jlm}\ n_j\ n_l\ n_m \nonumber
\end{eqnarray}
where $n_i$ are the number densities of the different species. 
The first and second right-hand terms present two-body reactions which yield losses 
(negative sign) and gains (positive sign) for species density $n_i$
with rates  $k_{2,ij}$ and $k_{2,jl}$, respectively. 
Three-body reactions are analogously accounted for by the third and fourth term 
with rates $k_{3,ijl}$ and $k_{3,jlm}$.
The resulting ordinary differential equation Eq.~\ref{eq:dndt} is of first order 
and has to be imposed for each chemical species separately. 
The whole problem thus requires the solution of a system of differential equations. 

Chemical reactions can have rates that differ by many orders of magnitude. 
Hence, not only the number densities of different species but also their temporal 
derivatives cover a large range, causing the system of equations to be stiff. 
Consequently, an implicit scheme should be used for the numerical solution of the 
problem.

\subsection{Time-dependent solution}
\label{sec:methodtd}

The time-dependent treatment of a chemical reaction network has been added 
as an optional separate computational step to the radiation 
hydrodynamics code \textsf{\mbox{CO$^5$BOLD}}. 
The general properties of the code are described in \citet{cobold} and 
\citet[][ hereafter W04]{wedemeyer04a}.
Following the general concept of operator splitting, 
the chemistry calculations are performed after the hydrodynamics step for each 
computational time step. The treatment of advection and chemical reactions 
is thus separated into subsequent steps. 
First, the number densities of all chemical species are advected with the hydrodynamic flow field. 
This is done analogously to the gas density within the applied Roe solver.
After the subsequent viscosity step the chemical reactions are handled for 
each grid cell separately, starting 
with the calculation of the reaction rates. Necessary input data are 
the local temperature, which is available as result of the hydrodynamics step, and 
the number densities of the involved chemical species of the previous 
time step. 
The basic rate is given by    
\begin{equation}                 
  \label{eq:ratebasic}
  k\ =\ \alpha\ {T_{300}}^{\beta}\  e^{-\gamma/T}\enspace,  
\end{equation}
where $T_{300} = T/300$~K. 
For catalytic reactions which involve a representative metal 
also the number density $n_\mathrm{M}$ of the metal enters:
\begin{equation}                 
  \label{eq:ratemetal}
  k\ = n_\mathrm{M}\ \alpha\ {T_{300}}^{\beta}\  e^{-\gamma/T} \enspace.
\end{equation}
Given the rates, the system of differential equations is defined
and then solved with an implicit BDF 
(backward differentiation formula) 
method. Here, the DVODE package is used
\citep{brown89}. The solver uses an internal computational time step 
which is adjusted automatically. 
The solution provides the number densities of the involved species 
after the overall computational time step prescribed by the foregoing hydrodynamics.

The radiative transfer is treated in a subsequent step frequency-independently 
(grey) and under strict assumption of local thermodynamic equilibrium (LTE). 
See W04 for more details. 

The boundary conditions are consistent with the hydrodynamics part of the code, 
i.e. in the presented simulations chemical species are advected across the lateral 
periodic boundaries and can leave the computational domain at top or bottom.
Like for the other boundary conditions, the number densities in the bottom grid layer 
are copied into the ghost cells below but 
are scaled to the mass density $\rho$ of those cells: 
\begin{equation}
\frac{\partial ( n_i / \rho )}{\partial z} = 0 
\quad
\Leftrightarrow 
\quad
n_{i}^\mathrm{ghost} = n_{i}^\mathrm{bottom} \, \rho^\mathrm{ghost}\, /\, \rho^\mathrm{bottom}
\enspace.
\end{equation}
This way the total particle numbers of the involved elements 
(here \element{H}, \element{C}, \element{O}, and a representative metal M) 
are almost perfectly conserved, except for a  decrease of typically $5\,10^{-4}$ per 
simulation hour in relative number due to mass loss at the upper 
boundary. 

\subsection{Chemical equilibrium}
\label{sec:methodce}

Next to the full time-dependent simulations, separate calculations have been performed 
in order to derive CO equilibrium densities. In this case, which is referred to as CE 
in the following, different snapshots of the time-dependent simulation sequence 
are used as initial condition with the number densities derived from the gas 
density in the same way as done for the start model (see Sect.~\ref{sec:model}). 
For each grid cell the pure chemistry calculations are performed analogously to the 
time-dependent treatment described in Sect.~\ref{sec:methodtd}, but with the 
local gas temperature kept constant and without advection, 
until the CO number density reaches an equilibrium value $n_\mathrm{CO, eq}$. 
Starting from an initial value $n_{\mathrm{CO}, 0} = n_\mathrm{CO} (t=0)$, the 
resulting temporal evolution of the CO number density follows
in most cases perfectly the function 
\begin{equation}
n_\mathrm{CO} (t) = ( n_{\mathrm{CO}, 0} - n_\mathrm{CO, eq} )\;  e^{-t / t_\mathrm{chem}}\; + n_{\mathrm{CO}, 0}\enspace, 
\end{equation}
which is the typical solution for the differential chemistry equation (Eq.~\ref{eq:dndt}).
The derived profile $n_\mathrm{CO} (t)$ then allows to determine the equilibrium value 
\begin{equation}
n_\mathrm{CO, CE} \equiv n_\mathrm{CO, eq} = \lim_{t\to\infty}  n_\mathrm{CO} (t)
\end{equation}
and the corresponding chemical timescale $t_\mathrm{chem}$ for every grid cell. 
In principle the assumption of instantaneous chemical equilibrium (ICE) should 
provide the same values, 
although that approach utilises chemical equilibrium constants instead of a 
reaction network.

For comparison, ICE number densities are calculated with the spectrum 
synthesis code RH (``Rybicki \& Hummer'') by \citet{uitenbroek01} for a number of snapshots 
from our time-dependent 2D model in the same way as done by \citet{uitenbroek00a}. 
The results of these calculations are here referred to as UICE.

\subsection{ICE and spectrum synthesis}
\label{sec:rhcode}

The RH~code (see Sect.~\ref{sec:methodce}) is also used for calculating spectra.  
We have modified this code, which by default computes the instantaneous 
chemical equilibrium densities for a user-defined set of molecules and their 
constituent atoms, to also accept the CO densities computed with 
 \textsf{\mbox{CO$^5$BOLD}} as input. 
The distribution over the various energy levels of the CO molecule is assumed 
to be in LTE, which has been shown to be a sound approximation \citep{ayres89}, 
so that the CO line opacities and line source functions are in LTE. 
We compute intensities in the 2142 -- 2145 cm$^{-1}$ range 
(4.662 -- 4.668~$\mu$m), which is observable from the ground and which 
contains CO lines with widely varying properties \citep{goorvitch94}. 
Due to the non-negligible contribution of Thomson scattering to the 
total opacity at the wavelengths studied, we have to lambda-iterate 
the angle-averaged radiation field $J_\nu$ to convergence. 
For those computations we use the A4 angular quadrature of \citet{carlson63}, 
which has 3 rays per octant. 
For the intensities in the vertical direction a separate formal solution 
was subsequently performed using the angle-averaged radiation field 
$J_\nu$ from the full solution to account for the scattering contributions.

Since scattering off free electrons is an important source of opacity in 
higher layers of the solar atmosphere, it is necessary to use realistic 
electron densities. 
The default LTE values, obtained by solving the Saha-Boltzmann equations for 
all species, are definitely unrealistic in the chromosphere since there 
the degree of ionisation is largely decoupled from the local temperature. 
Instead of the default LTE values we derive electron densities from the FAL-C model \citep{fal91} 
by means of interpolation of the $N_{\rm e}(N_{\rm H})$ dependence. 
   
In order to be consistent with the chemo-hydrodynamic calculations
in this paper, the same chemical abundances have been adopted 
(see Sect.~\ref{sec:model}). 
By default, the following molecules are taken into account in RH: 
H$_2$, H$_2$$^+$, C$_2$, N$_2$, O$_2$, CH, CO, CN, NH, NO, OH, and H$_2$O.
While this set is used for the spectrum synthesis, we did also a 
calculation with 
a reduced set comprising only the molecules that  
are present in the chemical reaction network (see Sect.~\ref{sec:network}). 
The results are discussed in Sect.~\ref{sec:discus}.

\section{Chemical reaction network}
\label{sec:network}

\begin{figure}[t] 
\centering 
  \resizebox{\hsize}{!}{\includegraphics{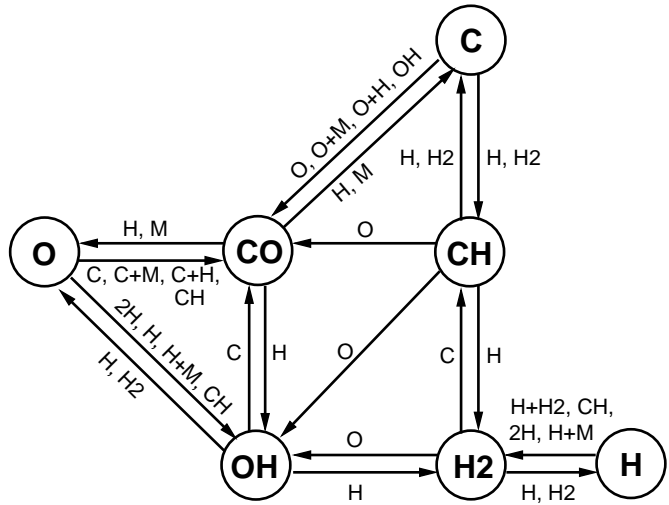}} 
  \caption{Chemical reaction network. The seven chemical species (not including 
  the representative metal) are connected via reactions which are listed in 
  Table~\ref{tab.reaction}. The 
  involved chemical species of the reactions are printed at the arrows.
  } 
  \label{fig.chemnetwork} 
\end{figure} 

\begin{table*}[t]
\caption[h]{Chemical reaction network: 
Involved are the species H, H$_2$, C, O, CO, CH, OH, a representative metal (M), and 
photons ($\nu$). 
The corresponding reaction rates are parameterised by the coefficients  $\alpha$, $\beta$, and $\gamma$, 
following the UMIST standard (see Eqs.~\ref{eq:ratebasic}-\ref{eq:ratemetal}).
The dimension of $\alpha$ is $\mathrm{cm}^3\mathrm{s}^{-1}$ and $\mathrm{cm}^6\mathrm{s}^{-1}$ 
for two- and three-body reactions, respectively. $\beta$ is dimensionless, whereas $\gamma$ is given 
in units of K.
The last column gives references to the origin of the rates: 
UMIST: \citet{leteuff2000}, 
KCD: Konnov's combustion database  \citep{konnov00}, 
W80:  \citet{westley80}
BDHL72: \citet{baulch72}
BDDG76: \citet{baulch76}; 
The ID numbers indicate where the reactions have been found first, regardless of the 
final source of the coefficients which is given in the last column. 
An ID number less than 5000 refers to UMIST, whereas numbers 5000 and above are 
assigned to additional reactions found in Konnov's database. Reactions listed 
by \citet{ayres96} are classified with numbers of $\ge$~7000. 
} 
\label{tab.reaction}
\centering
\begin{tabular}[b]{r l c l rrr l}
\hline
ID nr. &\multicolumn{3}{l}{reaction}&$\alpha$ & $\beta$ & $\gamma$           &ref.\\
 &\multicolumn{3}{l}{        }&        &         & $[\mathrm{K}]$     &    \\
\hline
\ \\[-3mm]
& \multicolumn{3}{c}{radiative association}& [cm$^3$~s$^{-1}$] & & & \\
\hline
3681&       H $+$       C           &$\rightarrow$&      CH $+$   $\nu$           &$1.00(-17)$&$    0.00$&$     0.0$& UMIST\\
3683&       H $+$       O           &$\rightarrow$&      OH $+$   $\nu$           &$9.90(-19)$&$   -0.38$&$     0.0$& UMIST\\
3707&       C $+$       O           &$\rightarrow$&      CO $+$   $\nu$           &$1.58(-17)$&$    0.34$&$  1297.4$& UMIST\\
\hline
\ \\[-3mm]
& \multicolumn{3}{c}{3-body association}& [cm$^6$~s$^{-1}$] & & & \\
\hline
5001&       H $+$       H $+$   H$_2$ &$\rightarrow$&   H$_2$ $+$   H$_2$           &$9.00(-33)$&$   -0.60$&$     0.0$& KCD\\
5002&       H $+$       H $+$       H &$\rightarrow$&   H$_2$ $+$       H           &$4.43(-28)$&$   -4.00$&$     0.0$& BDHL72\\
7000&       O $+$       H $+$       H &$\rightarrow$&      OH $+$       H           &$1.00(-32)$&$    0.00$&$     0.0$& BDHL72\\
7001&       C $+$       O $+$       H &$\rightarrow$&      CO $+$       H           &$2.14(-29)$&$   -3.08$&$ -2114.0$& BDDG76\\
\hline
\ \\[-3mm]
& \multicolumn{3}{c}{Species exchange}& [cm$^3$~s$^{-1}$] & & & \\
\hline
   1&       H $+$      CH            &$\rightarrow$&       C $+$   H$_2$           &$2.70(-11)$&$    0.38$&$     0.0$& UMIST\\
   8&       H $+$      OH            &$\rightarrow$&       O $+$   H$_2$           &$6.99(-14)$&$    2.80$&$  1950.0$& UMIST\\
  14&       H $+$      CO            &$\rightarrow$&      OH $+$       C           &$5.75(-10)$&$    0.50$&$ 77755.0$& W80\\
  42&   H$_2$ $+$       C            &$\rightarrow$&      CH $+$       H           &$6.64(-10)$&$    0.00$&$ 11700.0$& UMIST\\
  48&   H$_2$ $+$       O            &$\rightarrow$&      OH $+$       H           &$3.14(-13)$&$    2.70$&$  3150.0$& UMIST\\
  66&       C $+$      OH            &$\rightarrow$&       O $+$      CH           &$2.25(-11)$&$    0.50$&$ 14800.0$& UMIST\\
  67&       C $+$      OH            &$\rightarrow$&      CO $+$       H           &$1.81(-11)$&$    0.50$&$     0.0$& W80\\
 102&      CH $+$       O            &$\rightarrow$&      OH $+$       C           &$2.52(-11)$&$    0.00$&$  2381.0$& UMIST\\
 104&      CH $+$       O            &$\rightarrow$&      CO $+$       H           &$1.02(-10)$&$    0.00$&$   914.0$& UMIST\\
\hline
\ \\[-3mm]
& \multicolumn{3}{c}{collisional dissociation}& [cm$^3$~s$^{-1}$] & & & \\
\hline
4060&       H $+$   H$_2$           &$\rightarrow$&       H $+$       H$+$       H&$4.67(-07)$&$   -1.00$&$ 55000.0$& UMIST\\
4061&       H $+$      CH           &$\rightarrow$&       C $+$       H$+$       H&$6.00(-09)$&$    0.00$&$ 40200.0$& UMIST\\
4062&       H $+$      OH           &$\rightarrow$&       O $+$       H$+$       H&$6.00(-09)$&$    0.00$&$ 50900.0$& UMIST\\
4069&   H$_2$ $+$   H$_2$           &$\rightarrow$&   H$_2$ $+$       H$+$       H&$1.00(-08)$&$    0.00$&$ 84100.0$& UMIST\\
4070&   H$_2$ $+$      CH           &$\rightarrow$&       C $+$   H$_2$$+$       H&$6.00(-09)$&$    0.00$&$ 40200.0$& UMIST\\
4071&   H$_2$ $+$      OH           &$\rightarrow$&       O $+$   H$_2$$+$       H&$6.00(-09)$&$    0.00$&$ 50900.0$& UMIST\\
7002&      CO $+$       H           &$\rightarrow$&       C $+$       O$+$       H&$2.79(-03)$&$   -3.52$&$128700.0$& BDDG76\\
\hline
\ \\[-3mm]
& \multicolumn{3}{c}{collision induced dissociation}& [cm$^3$~s$^{-1}$] & & & \\
\hline
4076&      CO $+$       M           &$\rightarrow$&       O $+$       C$+$       M&$2.79(-03)$&$   -3.52$&$128700.0$& BDDG76\\
\hline
\ \\[-3mm]
& \multicolumn{3}{c}{catalysed termolecular reactions}& [cm$^6$~s$^{-1}$] & & & \\
\hline
4079&       H $+$       M $+$       O&$\rightarrow$&      OH $+$       M           &$4.33(-32)$&$   -1.00$&$     0.0$& UMIST\\
5000&       H $+$       M $+$       H&$\rightarrow$&   H$_2$ $+$       M           &$6.43(-33)$&$   -1.00$&$     0.0$& KCD\\
4097&       C $+$       M $+$       O&$\rightarrow$&      CO $+$       M           &$2.14(-29)$&$   -3.08$&$ -2114.0$& BDDG76\\
\hline
\end{tabular}
\end{table*}

In the present study we take into account eight chemical species: 
H, H$_2$, C, O, CO, CH, OH, and M, where M stands for a representative inert catalyst 
(also shortly referred to as 'metal'). 
For the standard model (see Sect.~\ref{sec:model}) the 
abundance of the representative metal was set to the one of helium.
Strictly speaking He is not a metal but the most abundant element that 
can be chosen as representative catalytic element M. 
This way an upper limit for the influence of M is provided.  
The species are connected via 27 reactions (see Table~\ref{tab.reaction}).
Ion-molecule reactions and photoreactions are excluded because, following \citet{asensio03},  
their influence on the total CO~concentration is negligible for heights of 
$\le 1000$~km in the solar atmosphere. 
The chemical input data are taken from different sources
\footnote{Note that the here given reference BDHL72 is apparently identical 
with BDDG72 as used by AR96.} 
which are discussed below. 

First, the UMIST database rate file {\em rate99} \citep{leteuff2000} has 
been reduced to reactions involving chemical species stated above. 

In addition, a few termolecular\footnote{termolecular: A reaction involving the simultaneous collision of three particles.} 
reactions are taken from the combustion 
database of \citet[][ \texttt{http://homepages.vub.ac.be/$\sim$akonnov}]{konnov00}. 
\citet{asensio03}, for example, based their chemical network entirely on the rates 
from this database. 

The list of reactions is further supplemented with three reactions present in  
Table~2 by \citet{ayres96} 
that are not found in UMIST (\#7000, \#7001, \#7002).
Note that their data are explicitely prepared for a gas temperature of 5000~K
because of the authors' interest in the CO reformation timescales in a previously 
molecule-free gas cooling below the H$^{-}$ equilibrium temperature. 

Some reactions are available from different sources and their reaction rate
coefficients differ noticeably. Some of the rates in Konnov's database for
instance are different from those in UMIST. They differ not only in 
the absolute value of the rate, but also in the type of reaction (temperature 
dependence, activation barrier). Some others agree surprisingly well. 
These ambiguities demand for a closer examination of the chemical input data. 
Here we present a examination of different available sources for reaction 
rates included in our chemical network (see Table~\ref{tab.reaction}). The different 
data are listed in Table~\ref{tab.reaccomp} and discussed below.

%
\begin{table*}[t]
\caption[h]{Comparison of chemical data from different sources. 
The reaction rates are parameterised by the coefficients  $\alpha$, $\beta$, and $\gamma$, 
following the UMIST standard. 
The dimension of $\alpha$ is $\mathrm{cm}^3\mathrm{s}^{-1}$ and $\mathrm{cm}^6\mathrm{s}^{-1}$ 
for two- and three-body reactions, respectively. $\beta$ is dimensionless, whereas $\gamma$ is given 
in units of K. The sixth column gives the temperature range in K over which the reaction rate is valid
and the seventh column its accuracy (following the UMIST error nomenclature, A: $<$~25\%, B: $<$~50\%,
C: within a factor 2, D: within an order of magnitude, E: highly uncertain).
The last column gives references to the origin of the rates: 
UMIST: \citet{leteuff2000}, 
AR96: \citet{ayres96}, 
KCD: Konnov's combustion database  \citep{konnov00}, 
W80:  \citet{westley80}
BDHL72: \citet{baulch72}
BDDG76: \citet{baulch76}
} 
\label{tab.reaccomp}
\centering
\begin{tabular}[h]{l l c l rrr cll}
\hline
ID nr.&\multicolumn{3}{l}{reaction}&$\alpha $&$\beta$&$\gamma$ & T range & acc. &ref.\\
\hline
1   & CH $+$ H     &$\rightarrow$& C  $+$ H$_2$   &$2.70(-11)$ &$ 0.38$ &$     0.0$ &  300 -- 2000 & B & UMIST (1)\\
    &              &                 &                &$1.31(-10)$ &$  0.0$ &$    80.5$ &           &   & KCD\\
\hline
8   &H   $+$ OH    &$\rightarrow$& O  $+$ H$_2$   &$6.99(-14)$ &$ 2.80$ &$  1950.0$ &  300 -- 2500 & A & UMIST (8)\\
    &              &                 &                &$3.35(-13)$ &$ 1.70$ &$     0.0$ &$\sim$5000 &   & AR96\\
\hline
14  &H   $+$ CO    &$\rightarrow$& OH $+$ C       &$1.10(-10)$ &$ 0.50$ &$ 77700.0$ &2590 -- 41000 & C & UMIST (14)\\
    &              &                 &                &$3.21(-35)$ &$16.20$ &$     0.0$ &$\sim$5000 &   & AR96\\
    &              &                 &                &$5.75(-10)$ &$ 0.50$ &$ 77755.0$ &typical T$_\mathrm{comb}$ &   & W80\\
\hline
48  & O  $+$ H$_2$ &$\rightarrow$& OH $+$ H       &$3.14(-13)$ &$  2.7$ &$  3150.0$ &  297 -- 3532 & A & UMIST (48)\\
    &              &                 &                &$2.86(-13)$ &$ 1.90$ &$     0.0$ &$\sim$5000 &   & AR96\\
    &              &                 &                &$4.10(-13)$ &$  2.7$ &$  3165.1$ &           &   & KCD\\
\hline
67  &  C $+$ OH    &$\rightarrow$& CO $+$ H       &$1.10(-10)$ &$  0.5$ &$     0.0$ &    10 -- 300 & C & UMIST (67)\\
    &              &                 &                &$1.22(-10)$ &$  0.5$ &$     0.0$ &$\sim$5000 &   & AR96\\
    &              &                 &                &$8.30(-11)$ &$  0.0$ &$     0.0$ &           &   & KCD\\
    &              &                 &                &$1.81(-11)$ &$  0.5$ &$     0.0$ &typical T$_\mathrm{comb}$ & & W80\\
\hline
102 & CH $+$ O     &$\rightarrow$& C  $+$ OH      &$2.52(-11)$ &$  0.0$ &$  2381.0$ &   10 -- 6000 & B & UMIST (102)\\
    &              &                 &                &$2.52(-11)$ &$  0.0$ &$  2380.1$ &           &   & KCD\\
\hline
103/4& CH $+$ O    &$\rightarrow$& CO $+$ H       &$6.60(-11)$ &$  0.0$ &$     0.0$ &   10 -- 2000 & A & UMIST (103)\\
    &              &                 &                &$1.02(-10)$ &$  0.0$ &$   914.0$ & 2000 -- 6000 & A & UMIST (104)\\
    &              &                 &                &$6.64(-11)$ &$  0.0$ &$     0.0$ &           &   & KCD\\
\hline
3683& H   $+$ O    &$\rightarrow$& OH $+$ $h\nu$  &$9.90(-19)$ &$-0.38$ &$     0.0$ &    10 -- 300 & C & UMIST (3683)\\
    &              &                 &                &$9.24(-19)$ &$-0.40$ &$     0.0$ &$\sim$5000 &   & AR96\\
\hline
3707& C   $+$ O    &$\rightarrow$& CO $+$ $h\nu$  &$1.58(-17)$ &$ 0.34$ &$  1297.4$ & 300 -- 13900 & B & UMIST (3707)\\
    &              &                 &                &$5.55(-18)$ &$ 0.60$ &$     0.0$ &$\sim$5000 &   & AR96\\
\hline
4060& H  $+$ H$_2$ &$\rightarrow$& H $+$ H $+$ H &$4.67(-07)$ &$-1.00$ &$ 55000.0$ &1833-41000 & C & UMIST (4060)\\
    &              &                 &                &$3.45(-21)$ &$ 6.60$ &$     0.0$ &$\sim$5000 &   & AR96\\
    &              &                 &                &$8.86(-04)$ &$-4.00$ &$ 51900.0$ & 3400 -- 5000 & D & BDHL72\\
\hline
4062& H   $+$ OH   &$\rightarrow$& O $+$ H $+$ H  &$6.00(-09)$ &$ 0.00$ &$ 50900.0$ &1696-41000 & C & UMIST (4062)\\
    &              &                 &                &$3.92(-26)$ &$10.40$ &$     0.0$ &$\sim$5000 &   & AR96\\
\hline
4076& CO  $+$ M    &$\rightarrow$& O  $+$ C $+$ M &$2.86(-03)$ &$-3.52$ &$112700.0$ &2000 -- 10000 & B & UMIST (4076)\\
    &              &                 &                &$2.79(-03)$ &$-3.52$ &$128700.0$ &7000 -- 15000 & D & BDDG76\\
\hline
5000& H $+$ H $+$ M&$\rightarrow$& H$_2$ $+$ M    &$6.43(-33)$ &$-1.0 $ &$     0.0$ &           &   & KCD\\
    &              &                 &                &$5.88(-33)$ &$-1.0 $ &$     0.0$ & 1700 -- 5000 & C & BDHL72\\
\hline
5001& H $+$ H $+$ H$_2$ &$\rightarrow$& H$_2$ $+$ H$_2$ &$9.00(-33)$ &$-0.6 $ &$0.0$ &           &   & KCD\\
    &                   &                 &                 &$2.39(-32)$ &$-1.0 $ &$0.0$ & 2500 -- 5000 & D & BDHL72\\

\hline
5002&H $+$ H $+$ H &$\rightarrow$& H$_2$ $+$ H    &$8.82(-33)$ &$ 0.0 $ &$     0.0$ &           &   & KCD\\
    &              &                 &                &$4.63(-28)$ &$-4.0 $ &$     0.0$ &$\sim$5000 &   & AR96\\
    &              &                 &                &$1.83(-31)$ &$-1.0 $ &$     0.0$ & & & P83\\
    &              &                 &                &$4.43(-28)$ &$-4.0 $ &$     0.0$ & 3400 -- 5000 & D & BDHL72\\
\hline
7000&O $+$ H $+$ H &$\rightarrow$& OH $+$ H       &$1.00(-32)$ &$  0.0$ &$     0.0$ &$\sim$5000 &   & AR96\\
    &              &                 &                &$2.76(-33)$ &$  0.0$ &$     0.0$ & low.limit & E & BDHL72\\
    &              &                 &                &$2.76(-32)$ &$  0.0$ &$     0.0$ & upp.limit & E & BDHL72\\
\hline
7001& C $+$ O $+$ H&$\rightarrow$& CO $+$ H       &$3.78(-29)$ &$-3.50$ &$     0.0$ &$\sim$5000 &   & AR96\\
    &              &                 &                &$2.14(-29)$ &$-3.08$ &$ -2114.0$ &7000 -- 14000 & D & BDDG76\\
\hline
7002& CO $+$ H     &$\rightarrow$& C $+$ O $+$ H  &$1.39(-46)$ &$22.80$ &$     0.0$ &$\sim$5000 &   & AR96\\
    &              &                 &                &$2.79(-03)$ &$-3.52$ &$128700.0$ &7000 -- 15000 & D & BDDG76\\
\hline
\end{tabular}
\end{table*}

\paragraph{Reactions 8, 48.}
Since Ayres \& Rabin fitted the {UMIST \citep[][ former version]{millar91} reaction rate explicitely 
for temperatures around 5000~K, their fit is poor for gas 
temperatures very different from that. 
The newer UMIST data \citep{leteuff2000} 
comes from the NIST database (\texttt{http://www.nist.gov}) and an accuracy of better than 25\% for
$300$~K~\mbox{$< T < 2500$~K}~(8) and $297$~K~\mbox{$< T < 3532$}~K~(48) is stated. 
At high temperatures, the real rates could be an order of magnitude larger 
than the UMIST rate. On the other hand, Konnov's rate for reaction \#48 is
very similar to the one of UMIST. 
The assumption that both are probably based on the same
experimental data supports our decision to use the UMIST reaction rate.

\paragraph{Reaction 14.}                                                            
Ayres \& Rabin derived this rate from the assumption of
detailed balance for reactions \#14 and \#67, where they calculated
the CO/OH equilibrium ratio from a Saha Ansatz. The high exponent $\beta$ 
indicates that this rate is only valid in a very narrow temperature
range around 5000~K. The UMIST rate is more than one order of magnitude smaller 
around 5000~K. We trace the UMIST rate back to \citet{mitchell84}
and further to \citet{westley80}. The original work of \citet{westley80}
gives a $\sim 5$ times higher rate than Mitchell and UMIST. Since the 
reason for this discrepancy is unknown, we use the original rate from
Westley.

\paragraph{Reaction 67.}
For this reaction, the source of the UMIST database is \citet{prasad80}.
The reaction rate did not change in UMIST between the 1991 and 1999 rate files. 
Konnov's rate is $\sim 5$ times 
smaller than that of UMIST around 5000~K and has no temperature dependence 
($8.30\,10^{-11}$ compared to $4.5\,10^{-10}$). \citet{westley80} quotes a 
rate of $7.39\,10^{-11}$ at 5000~K, a factor of 6 lower than UMIST. Both,
Konnov's rate and the one of Westley are valid for typical combustion 
temperatures, while the UMIST rate is only valid up to 300~K.

There are two reasons for choosing the \citet{westley80} rate: (1)
the reactions \#14 and \#67 form a pair of forward and backward reaction
and hence the same data should be used for them. (2) This reaction
rate is valid for typical combustion temperatures and hence more
appropriate to the temperature range covered in our solar atmosphere
simulations.

\paragraph{Reaction 3683.}                                                             
Ayres \& Rabin state UMIST as source for this rate coefficient although 
apparently they slightly altered it. For the present chemical network the original 
UMIST rate is used.

\paragraph{Reaction 3707.}
After detailed examination of the original data from \citet{dalgarno90}, 
and the fits by Ayres \& Rabin and UMIST, we adopted the rate coefficients provided by 
UMIST \citep{leteuff2000}. The latter give the best fit in the temperature 
range between 2000 and 8000~K.

\paragraph{Reaction 4060.}                                                            
Again, Ayres \& Rabin's fit of the rate by \citet{baulch72} holds around 
$T \sim 5000$~K. 
Outside that temperature range it systematically overestimates the original data 
by orders of magnitude. 
In contrast, the UMIST rate deviates by not more than a factor of 3 from the 
original data in the range 3400~K to 5000~K.
Even though the source of the UMIST fit is unknown to us, we favor it because 
the rate covers a larger temperature 
range and nicely follows the trend of the experimental data, which display 
a steeper gradient --- steeper than the \citet{baulch72} formula --- at 
temperatures below 3400~K. 
Baulch et al.\ state that their rate is only a 
tentative suggestion in a narrow temperature range.

\paragraph{Reaction 4062.}
The Ayres \& Rabin fit is again only valid for temperatures around 5000~K. 
We use the UMIST data instead, because it covers the whole temperature
range of interest.

\paragraph{Reaction 4076.}
The UMIST database refers to \citet{petuchowski89}, and from there to \citet{baulch76} 
(see rate \#7002). Apparently the exponent $\beta$ changed in the referencing process, 
the factor $\alpha$ suffered a rounding error and the error margin has been quoted 
wrongly. Hence, we stick to the original rate of \citet{baulch76}.

\paragraph{Reaction 5002.}
We cannot judge the accuracy of Konnov's rate over the temperature interval
$2000$~K~\mbox{$< T < 8000$}~K, but he refers back to \citet{cohen83}. \citet{palla83} refer 
back to a publication, where Cohen is a co-author \citep{jacobs67}. In this case,
\citet{cohen83} is newer and presents a literature review for rate constants. 
The difference in the rates by Konnov and \citet{baulch72} is the strong
temperature dependence of the rate in the latter reference. Since Baulch
et al. (1972) is a well established and documented database -- often used
by UMIST as well --, we use their rate constants. 

\paragraph{Reaction 7000.}
Owing to the lack of alternatives we adopt the rate by \citet{baulch72}
altough it should be regarded as a mere order of magnitude estimate. 

\paragraph{Reactions 7001, 7002.}                                                            
Ayres \& Rabin's rate is again restricted to the 5000~K temperature range.
The \citet{baulch76} fit assumes that H is very similar to Ar and it covers 
a larger temperature range, $7000$~K~\mbox{$< T < 15\,000$}~K. Hence, we use the latter
rate.

\section{Numerical model} 
\label{sec:model} 

The two-dimensional numerical model consists of 120 by 140 grid cells each with 
a horizontal width of 40~km. 
The height of the cells is 50~km at the bottom (at \mbox{$z = -1484$~km}) 
and smoothly decreases to 14~km for all heights above $z = -531$~km.  
The upper boundary is located at 1016~km. 
Note that the origin of the height axis is defined by the horizontal and temporal 
average of Rosseland optical depth unity ($\tau = 1$).
The total extent of the computational domain is thus 4800~km $\times$ 2500~km. 
Most numerical parameters are adopted from the recent 3D simulations by 
\citet{wedemeyer04a}, including the usage of grey OPAL-PHOENIX opacity
\citep{opal, hauschildt97}. 
The preliminary start model was extracted from the 3D model, too. In order to ensure 
relaxation the model was advanced several simulation hours before it was 
finally supplemented with arrays for number densities of the involved chemical species. 
For each grid cell the same 
constant chemical composition is assumed. The required abundances of the atomic species 
are set to $[\element{C}] = 8.39$ \citep{asplund05a}, and 
$[\element{O}] = 8.66$ \citep{asplund04a}, 
where $[\element{H}] \equiv 12$.
The initial abundances of the molecular species are set to $10^{-20}$ times the 
hydrogen density for CO, CH, and CO, and to $10^{-4}$ for \mbox{\element{H}$_2$}.  
The abundance of the representative metal is set equal to the helium abundance 
($[\element{He}] = 11.00$) and thus provides an upper limit for the influence of the 
metal. 
The initial number densities are then directly calculated from the gas density of the 
final start model for the constant chemical composition described above. 

For the standard model in total 86000~s have been calculated. 
The first 36000~s are reserved to ensure a sufficient chemical relaxation of the 
model and are therefore excluded from analysis.   
The global time step, which is relevant for hydrodynamics and radiative 
transfer, was typically 0.2~s -- 0.4~s, whereas the time step for the chemistry 
calculations has been adjusted by the solver itself for each individual 
cell and global time step. 

In addition to the standard model more simulations with altered chemical network have been 
produced in order to investigate the influence of individual reactions or 
reaction groups on the formation of CO. All simulations covered a time span 
of at least 7000~s.  
See Sect.~ \ref{sec:reacimport} for more details.

\section{Results} 
\label{sec:results} 

\begin{figure*}[pt] 
\centering 
  \includegraphics[width=170mm]{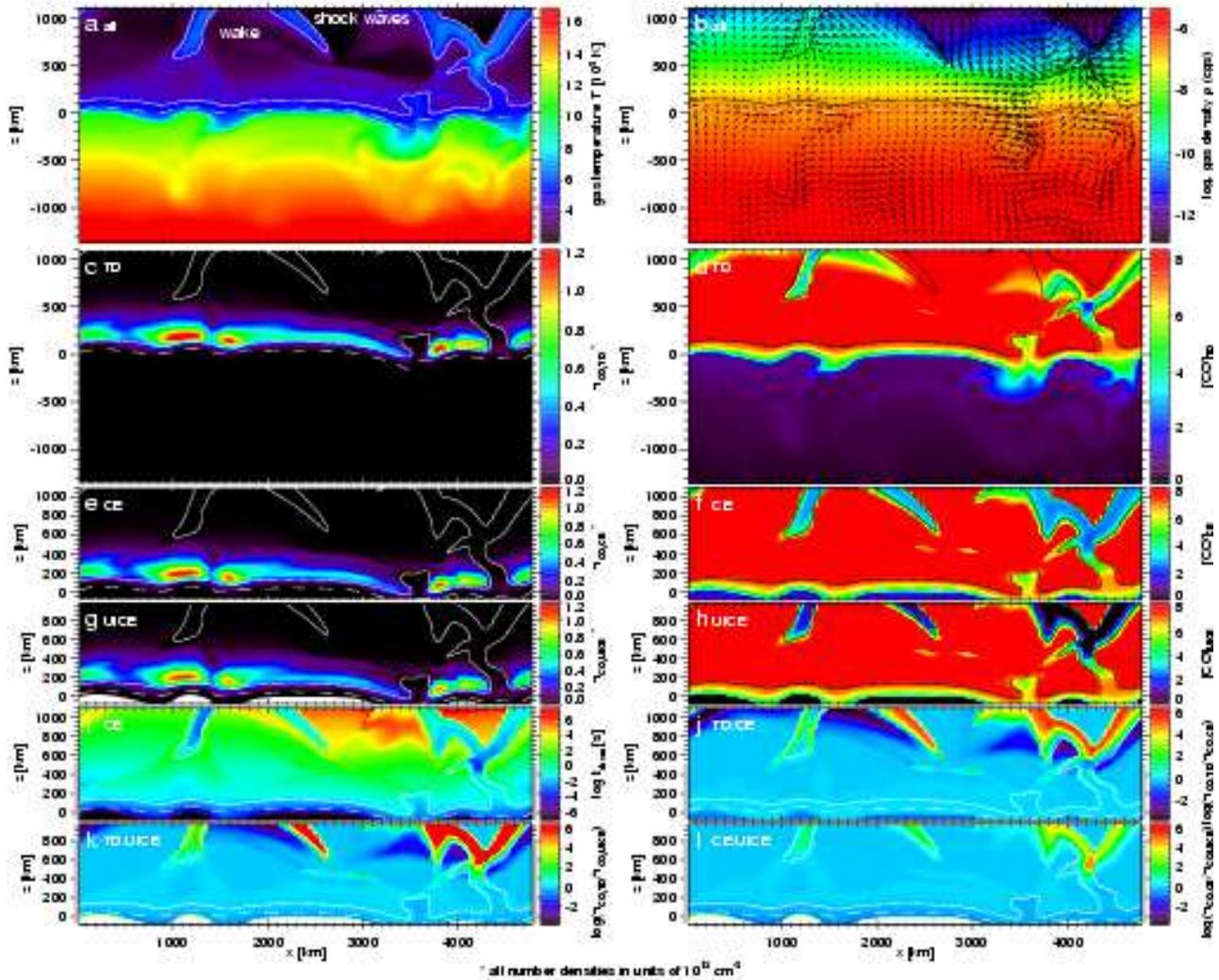}
  \caption{Single time step of the two-dimensional model ($t=70390$~s). 
  The upper row shows gas temperature $T$ (\textbf{a}) and 
  logarithmic gas density $\log \rho$ (\textbf{b}), including the velocity field
  which is represented by arrows. The length of the arrows corresponds to the 
  distance a fluid element travels within 15~s time (same scale as $x$, $z$-axes). 
  The resulting CO 
  number density $n_\mathrm{CO,TD}$ is presented in panel \textbf{c} and  
  additionally as abundance $[\mathrm{CO}]_\mathrm{TD}$ (\textbf{d})  
  on the usual logarithmic scale with [H]=12. 
  The third row presents the results of the equilibrium calculation (CE) for the same time 
  step (see text for more details): 
  CO number density  $n_\mathrm{CO,CE}$ (\textbf{e}) and 
  CO abundance $[\mathrm{CO}]_\mathrm{CE}$ (\textbf{f}). 
  In the next row (\textbf{g-h}) the corresponding results for the UICE calculation are displayed. 
  To enable direct comparisons of the different calculations, the data ranges
  are the same in the second to fourth row. 
  The chemical timescale, which has been derived from the CE simulation, is shown in panel~\textbf{i}.
  The differences between the three cases are plotted in terms of logarithmic fractions for the 
  pairs TD/CE, TD/UICE, and CE/UICE in panels \textbf{j}-\textbf{l}, respectively.
  In the lowest row all cells with temperatures above 9000~K appear white since 
  those are excluded in the UICE calculations. 
  The corresponding color/greyscale-coding is shown next to each panel. 
  The curves mark the height of average Rosseland optical depth unity (dashed) and 
  $T = 5000$~K (solid). 
  The solid curves clearly outline propagating shock waves in the model chromosphere.
  Note that for the cases CE and UICE only the upper part of the model is displayed 
  whereas the whole height extent is shown in the upper panels for the time-dependent 
  simulation.
  } 
  \label{fig.xzslices} 
\end{figure*} 

\subsection{Two-dimensional distribution}
\label{sec:spatdist}

In Fig.~\ref{fig.xzslices} a snapshot of the 2D model after a 
simulation time of 70390~s is presented. Each panel displays a different quantity as 
function of horizontal ($x$) and  vertical ($z$) position
for the time-dependent simulation (TD) but also for the equilibrium calculations 
CE and UICE (see Sect.~\ref{sec:methodce}), all based on the same time step .
The first panel displays the gas temperature, exhibiting a few granules and 
intergranular downflows. In contrast to positions above  granule interiors the 
temperature is increased above the downflows.  
These high-temperature regions in the middle photosphere, which 
form the reversed granulation pattern when seen from above 
\citep[see ][ W04]{leenaarts05}, 
are due to compression heating as result of convective overshoot. 
There are also high-temperature regions in the  upper layers which are produced by 
propagating shock waves (see W04). 
An example for the latter can be seen in the upper part of panels a and b, outlined 
by the contour line for $T = 5000$~K.
Both phenomena induce spatial and temporal inhomogeneities of the thermal structure and thus 
provide important constraints on the CO distribution. 
For clarity a contour line for $T = 5000$~K is drawn (white line).
The horizontal variations appear much smaller in terms of 
logarithmic gas density (panel~b) which on the other hand exhibits a  
decrease by several orders of magnitude from bottom to top of the model. 

The structuring in temperature and density has its direct influence on the 
CO number density $n_\mathrm{CO, TD}$ in panel~c which is alternatively presented as CO abundance (panel~d).  
CO is located mostly within a thin layer in the middle photosphere and 
closely follows changes of the gas temperature (see panel a), 
filling the cool interiors of the reversed granulation pattern.
Due to the steep density gradient in the atmosphere the CO number density decreases rapidly 
with height. Nevertheless, the relative number of CO molecules remains high in the 
upper layers as can be seen from the abundance [CO], which is defined as the ratio of 
CO particles and total number of hydrogen atoms (including those bound in molecules) on the commonly 
used logarithmic scale with [H]$\equiv$12. 
Here, [CO] is typically close to a value of 8.3 but it is strongly reduced at positions where 
the gas temperature exceeds $\sim 5000$~K (indicated by the white lines). 
This is true for the convection zone below the photosphere but also for the propagating 
shock waves which are a ubiquitous phenomenon in the upper layers. For the latter 
the abundance reveals a shift of the correlation between temperature and CO density which 
is otherwise strong in the layers below. 
[CO] is still high at the fronts of the  displayed shock waves, but it is not restored immediately in the wake. 
This is clear evidence that CO does not react instantaneously to thermal changes but rather 
on a chemical timescale 
(shown in panel~i, see Sects.~\ref{sec:rhot} and \ref{sec:chemtimescale} for more details) 
which is longer than the dynamic timescale due to the wave propagation.  
The dynamic behaviour and in particular the almost instantaneous reaction
of the CO number density to temperature changes in the photosphere
can be seen in an animation provided as online material\footnote{
Additional material can be found at \\
\texttt{http://www.kis.uni-freiburg.de/$\sim$sven/research/co.html}.}.

\begin{figure}[tp] 
\centering 
  \includegraphics[width=7.5cm]{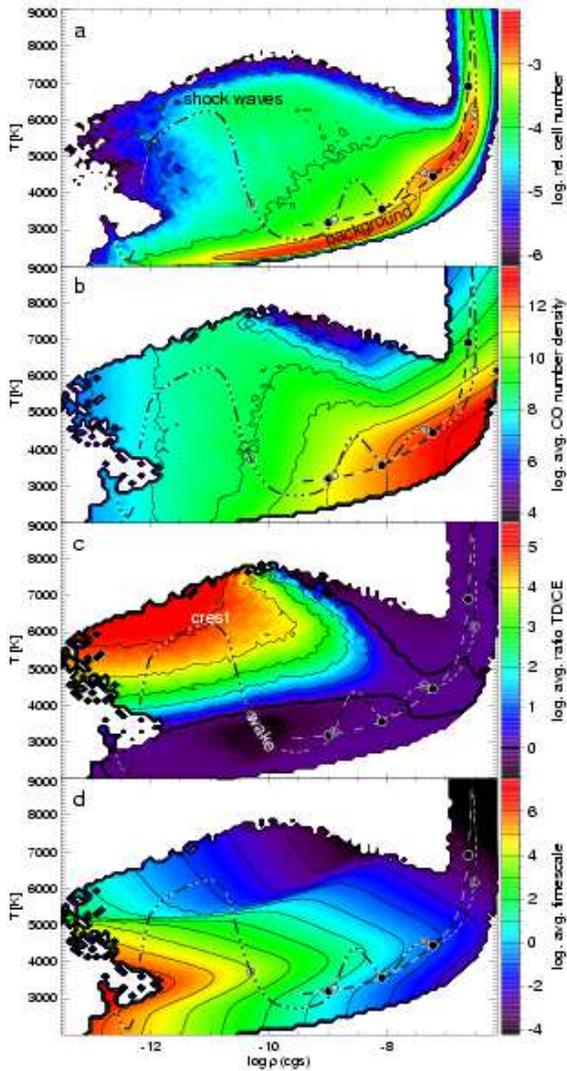}
  \caption{Quantites as function of logarithmic gas density and gas temperature
   for a sample of 139 snapshots:
  \textbf{a)} log. relative number of cells,
  \textbf{b)} log. averaged CO number density of TD case ($\log\,\langle n_\mathrm{CO,TD} \rangle$),
  \textbf{c)} log. averaged ratio of TD and CE cases ($\log\,\langle n_\mathrm{CO,TD}/n_\mathrm{CO,CE} \rangle$),
  \textbf{d)} log. averaged chemical timescale (CE);
  Contours and colors/grey scales are explained in the legend right next to 
  each panel. 
  The thick dashed line represents the horizontally and temporally averaged 
  stratification of the whole sequence, whereas the thick triple-dot-dashed line 
  marks a vertical column in the representative snapshot (see Fig.~\ref{fig.xzslices})
  at $x = 2380$~km, crossing a prominent shock wave. 
  The circles mark heights of $z = 250$~km, $500$~km, and $750$~km, respectively. 
  } 
  \label{fig.rhot} 
\end{figure} 

The results of the corresponding equilibrium calculations CE and UICE 
are shown in panels e-f, and g-h, respectively. 
The logarithmic ratios of the CO number densities of the different cases can be seen in panels j-l 
(j: TD/CE, k: TD/UICE, l: CE/UICE). 
Note that for UICE temperatures greater than 9000~K automatically result in a zero 
CO number density. Those cells are blanked in the plots.

The CE case is very similar to the time-dependent result almost everywhere in the photosphere 
but differs strongly in chromospheric shock waves and their wakes. 
Although the [CO] abundance is strongly reduced directly at the wave crest in both cases,  
the time-dependent calculations still result in a higher density at the front of the 
propating wave and in a lower one in the wake. 
This effect is caused by the essential but wrong assumption in CE 
that chemical equilibrium is reached 
before the gas temperature can change, analogous to the ICE assumption. 
Hence, CO equilibrium number density and local gas temperature do correlate strongly here, 
in contrast to the time-dependent simulation which does take into account finite chemical 
timescales in combination with the fast propagation of thermal inhomogeneities. 
In the latter (more realistic) case CO is 
only gradually dissociated when a hot wave arrives and does not form again instantaneously 
after the passage.
The equilibrium approach by nature cannot take into account this 
behaviour. 

The case  UICE yields a very similar picture but with somewhat smaller CO number density. 
Similar to CE, the deviations from the time-dependent model are small in the 
photosphere but become significant in the chromospheric shock waves. 
Finally, we compare CE and UICE: Deviations are very small in the photosphere but 
still get large at high temperatures in the centres of chromospheric shock waves. 
This remaining discrepancy must be attributed to differences 
in chemical input data for which equilibrium constants are used for UICE but the  
chemical reaction network for CE. 
In particular the reaction rate coefficients must be considered as potential 
source of errors since they are mostly defined for very limited temperature ranges 
only (see Table~\ref{tab.reaccomp}). At the crests of shock waves, where the largest 
discrepancies are found, the temperatures are high and thus exceed the 
stated range for many reactions.

\subsection{Dependence on gas density and temperature}
\label{sec:rhot}

For a sample of those 139 snapshots that are 
available both for the TD and the CE calculations 
different quantities are calculated as function of gas density and temperature. 
All grid cells in this sample -- regardless of their spatial and temporal 
position -- are sorted into discrete bins for logarithmic gas density $\log \rho$ and gas 
temperature $T$. 
The number of matching cells for each bin (Fig.~\ref{fig.rhot}a) is highest 
close to the average stratification (see dashed line) but much smaller for small 
densities and high temperatures in the upper left part of the plot which 
represents the domain of chromospheric shock waves. Furthermore, a gradual 
bifurcation can be seen with a cool background close to the average stratification 
and a hot component due to shock waves above (see also W04). 
The upper right part of the distribution corresponds to the top of the 
convection zone.

Panel~b reveals the clear tendency of CO being concentrated 
where the temperatures are low and gas densities are high. 
These conditions are best realised in the 
cooler regions above granule interiors in the low and middle photosphere of 
the model atmosphere. 
This causes the absolute number density to correlate best with a temperature of 
$\sim 4400$~K. 

More remarkable is the difference between TD and CE, which exhibits a 
significant pattern in panel~c. While the ratio of the corresponding CO 
number densities is small close to the 
average strafication, large values are the rule in the hot shock wave domain. 
The deviations in the wakes, which are due to a finite timescale on which 
the afore dissociated CO builds up again, show up as an ``eye'' 
around $\log \rho \sim -10$ and $T \sim 3000$~K.  
The triple-dot-dashed line represents a column in the exemplary snapshot at 
$x = 2380$~km (see Fig.~\ref{fig.xzslices}) which crosses a prominent shock 
wave. Consequently, this line traverses the ``eye'' and the shock wave domain 
in the ($\log \rho$, $T$)-plot.

Finally, the chemical timescale (Fig.~\ref{fig.rhot}d) derived in the CE calculations 
exhibits basically three  
things: First, the timescale grows strongly with decreasing gas density. 
Second, the skew in the shock wave domain indicates that there the 
timescales are much shorter due to a reduced final equilibrium concentration. 
And third, at a given gas density and with that at a given 
geometrical height a large range of values is present, 
making it difficult to define a meaningful average (see also Sect.~\ref{sec:chemtimescale}).

\subsection{Height distribution}

\begin{figure}[t] 
\centering 
  \resizebox{\hsize}{!}{\includegraphics{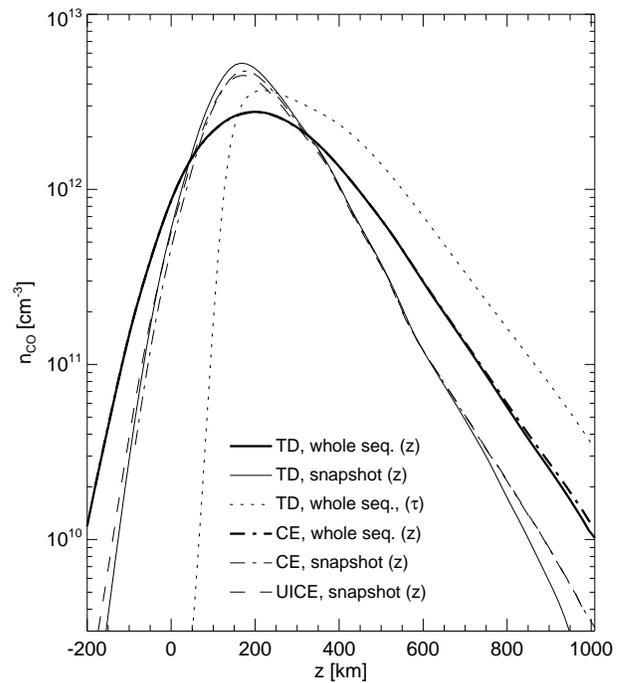}} 
  \caption{Average CO number density as function of geometrical 
  height for the whole time-dependent simulation sequence 
  (thick solid) and for the exemplary time step shown 
  in Fig.~\ref{fig.xzslices} (thin solid). 
  Also the results of the equilibrium calculations (CE) are displayed for 
  the whole sequence (thick dot-dashed) and for the exemplary 
  time step (thin dot-dashed), and for UICE calculation (thin dashed, snapshot only).
  The dotted line represents the average for the whole sequence, too, but 
  averaged on the Rosseland optical depth scale. 
  } 
  \label{fig.avgncoz} 
\end{figure} 

Averaging the CO number density horizontally and in time over a time span of 50000~s 
on the geometric height scale of the standard model results in the height distribution 
displayed in Fig.~\ref{fig.avgncoz}. 
It shows a prominent peak of 
\mbox{$\max (< n_\mathrm{CO}(z) >_{x,t}) = 2.81\, 10^{12}\,\mathrm{cm}^{-3}$}
at a height of \mbox{$z = 197$~km}.
Note that averaging on the Rosseland optical depth scale leads to a somewhat higher maximum of 
\mbox{$\max (< n_\mathrm{CO}(\tau) >_{x,t}) = 3.68\,10^{12}\,\mathrm{cm}^{-3}$} at 
an optical depth $\log \tau = -1.17$ which on average corresponds to a geometrical 
height of \mbox{$z = 217$~km}. 
The corresponding average abundance increases rapidly in the photosphere until 
a value of $\sim 8.3$ is reached at the bottom of chromosphere where it stays 
roughly constant. 
For the exemplary time step, which is also shown in  Fig.~\ref{fig.avgncoz}, the 
horizontally averaged CO distribution is very close to the corresponding 
equilibrium values (CE) but starts to deviate increasingly above the photosphere 
(see thin solid and dot-dashed lines). 
This result is expected since deviations mostly occur in the hot shock waves which 
become prominent usually in the low chromosphere and above 
(see Fig.~\ref{fig.xzslices}j). 
In the photosphere the conditions are right for equilibrium.
The same behaviour is found for the UICE case which perfectly matches the average 
CE profile for the examplary time step. However, the maximum is smaller by a factor of 0.86 
in the UICE case compared to TD and CE (see Fig.~\ref{fig.avgncoz}). 


\subsection{Chemical timescales}
\label{sec:chemtimescale}

Already the fact that the CO number density closely outlines the hotter regions 
in the low and middle photosphere (see Fig.~\ref{fig.xzslices}) indicates that 
there the chemical timescales are small compared to the timescales on which 
the atmospheric conditions change. 
Strictly speaking, chemical equilibrium refers to 
a given thermodynamic state of the atmosphere, for instance expressed in terms of 
gas temperature and particle densities. 
For reaching equilibrium it is thus essential that these quantities do stay 
constant. But under continuously varying conditions the chemical equilibrium can only 
be realised if the corresponding timescale is shorter than the dynamic one. 
Otherwise the concept of chemical timescales looses its meaning. 

Nevertheless, chemical time scales are in general very helpful since they allow to 
estimate the importance of individual reactions or reaction groups and, furthermore, 
are needed to decide whether the assumption of instantaneous chemical 
equilibrium is valid or time-dependent simulations are necessary. 

In principle the chemical time scale of a reaction is given by its inverse rate  
\citep[see, e.g.,][]{herbst73,ayres96}, for instance, 
\begin{equation}
t_\mathrm{chem} = ( k_{7002}\ n_\mathrm {H} )^{-1}
\end{equation}
for the two-body CO dissociating reaction \#7002 (see Table~\ref{tab.reaction}).
The required equilibrium number densities of the involved species can
be estimated for individual reactions/reaction pairs in the way shown by 
\citet{herbst73} and \citet{ayres96} but are 
difficult to be obtained in the present case of a complex chemical reaction network. 
Already the assumption of equilibrium number densities for the involved species 
is questionable since those species undergo a chemical 
evolution themselves. Each reaction, which 
e.g. forms CO, does not only depend on the local gas temperature and thus on the 
corresponding rate coefficient $k$ but also on the number densities of the required 
reactants. If now one of the involved species is not available in sufficient amount, 
owing to another less efficient reaction producing this species, this ``bottle-neck''
reduces the total effectivity of this formation channel. A chemical timescale of an 
individual reaction is thus only of limited scope under non-equilibrium conditions and 
rather must be seen in the context of the whole reaction network. 

The CE calculations described in Sect.~\ref{sec:methodce} do take into account 
the chemical evolution of the whole reaction network and allow to derive 
total effective chemical timescales and corresponding CO equilibrium densities. 
For the exemplary snapshot the resulting CE timescales are shown in 
Fig.~\ref{fig.xzslices}i, while the dependence on logarithmic gas density and 
temperature is displayed in Fig.~\ref{fig.rhot}d. 
These two figures make clear that a high temperature results in a low final CO 
number density and a related short timescale, whereas low temperatures lead 
to long timescales. 
Hence, the strong thermal inhomogeneities of the model atmosphere 
result in a very large spread of timescales, covering many magnitudes, 
which makes it hard to define a representative average. 

\begin{figure}[t] 
\centering 
  \resizebox{\hsize}{!}{\includegraphics{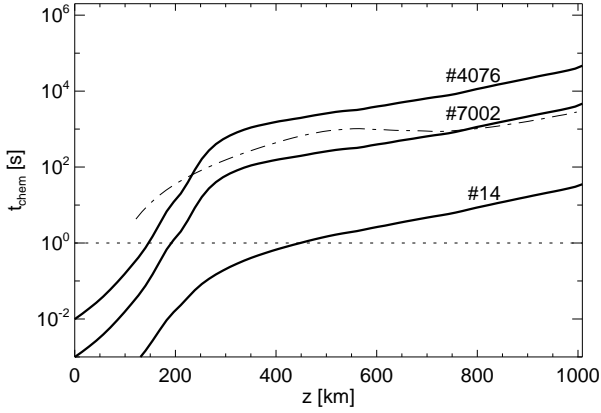}} 
  \caption{Chemical timescales  
   calculated from the average rates for the three dissociating reactions
   \#14, \#4076, and \#7002 (thick solid) and the result from the time-dependent 
   calculation in W04 (thin dot-dashed).
  } 
  \label{fig.timescale} 
\end{figure} 

Instead, we present average timescales for the three CO dissociating 
reactions \#14, \#4076, and \#7002 in Fig.~\ref{fig.timescale} in order to 
quantify the relative contribution of these reactions.  
For this the
reaction rates
are calculated for each grid cell for snapshots with a cadence of 100~s. 
After averaging temporally and horizontally for each reaction independently, the 
inverse average rates give the chemical timescales displayed in the figure. 
Obviously, the time scale for reaction \#14 is much shorter than those for the other two. 
Reaction \#7002 is typically two orders of magnitude slower, reaction \#4076 even around 
three. Consequently the total chemical timescale, given by the inverse sum of the 
rates of the three destruction reactions, is governed by reaction 14 alone. 
At a height of $z =646$~km, where the average mass density is 
$\langle \rho \rangle_{x,t} = 10^{15}\,\mathrm{g\,cm}^{-3}$, the chemical timescale for reaction 
\#14 is only $\sim 3$~s which corresponds to a gas temperature of $\sim 4900$~K. 
This can be compared to $0.7$~s for the rate coefficients given by \citet{ayres96} 
(see Table~\ref{tab.reaccomp}). 
However, also for this particular column density a large range of temperatures and 
related timescales is found. 
Hence, the average numbers should be considered as rough estimates only. 

The timescale for reaction \#7002 deviates to some extent from the result in W04. 
This is mainly due to the fact that the reaction coefficients by \citet{ayres96} 
were used in W04 instead of those by \citet{baulch76} as done in this work.
The two parameterisations only lead to similar 
reaction rates close to 5000~K and deviate otherwise. 
Nevertheless the average timescales agree quite well above $z \sim 700$~km. 

\begin{figure*}[t] 
  \centering 
  \includegraphics{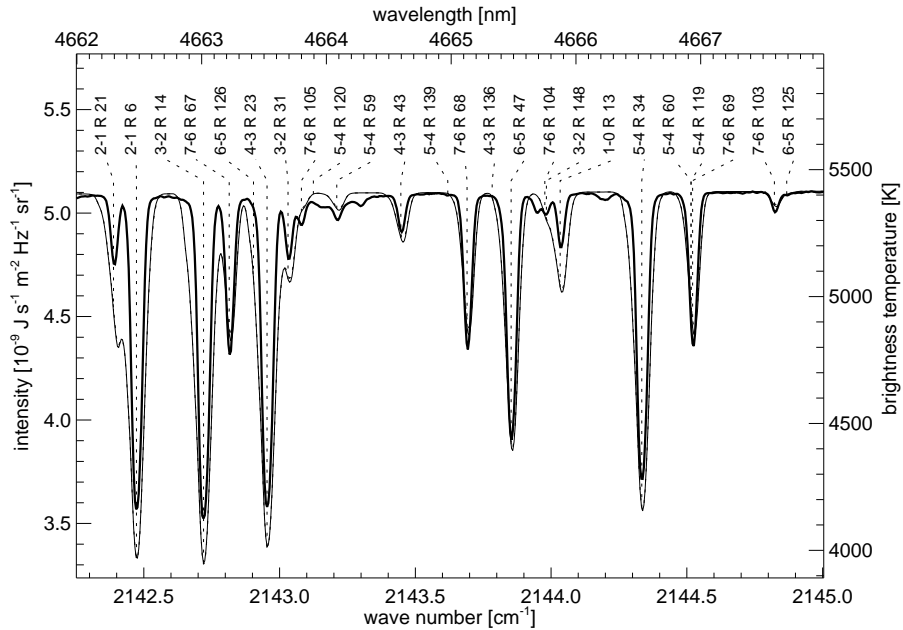}
  \caption{ Spectra near $\lambda = 4.7\,\mu\mathrm{m}$ with 
    fundamental vibration rotation lines of carbon monoxide (R~branch): 
     Average synthetic spectra with CO densities from time-dependent 2D model 
     (thin solid) compared to ATMOS3 data (thick solid).  
     Note that the UICE calculation exactly coincides with the time-dependent 
     case. 
  } 
  \label{fig.spectrum} 
\end{figure*} 

\subsection{Importance of individual reactions}
\label{sec:reacimport}

Next to the distribution of CO also the 
relative contribution of individual reactions is of great interest.  
This information allows to reduce the chemical 
reaction network as much as possible by excluding irrelevant reactions 
and thus increasing computational performance. 
One also gets to know which reactions are most important.

In our reaction network three reactions are dissociating carbon monoxide:\\

\begin{tabular}[h]{lllcll}
\#14  &:&\quad CO $+$ H &$\rightarrow$& C + OH&,\\
\#4076&:&\quad CO $+$ M &$\rightarrow$& C + O+ M&,\\
\#7002&:&\quad CO $+$ H &$\rightarrow$& C + O+ H&.\\
\end{tabular}\\

\noindent As we showed in Sect.~\ref{sec:chemtimescale} 
the species exchange reaction \#14 is by far the most efficient 
dissociation channel whereas \#7002 and \#4076 only contribute 
$10^{-2}$ and $10^{-3}$ in the low photosphere and $10^{-3}$ and $10^{-4}$
above, respectively, in terms of number of dissociated molecules per time. 

There are five reactions that form CO: \\

\begin{tabular}[h]{lllcll}
\#67  &:&\quad       C $+$      OH           &$\rightarrow$&      CO $+$       H&,\\           
\#104 &:&\quad       O $+$      CH           &$\rightarrow$&      CO $+$       H&,\\           
\#3707&:&\quad       C $+$       O           &$\rightarrow$&      CO $+$   $\nu$&,\\           
\#4097&:&\quad       C $+$       O $+$       M&$\rightarrow$&      CO $+$       M&,\\           
\#7001&:&\quad       C $+$       O $+$       H&$\rightarrow$&      CO $+$       H&.\\           
\end{tabular}\\

\noindent 
Again, the species exchange reactions are the most 
effective ones under the conditions of the solar atmosphere. 
These reactions, namely \#67 (via OH) and \#104 (via CH) have similar rates, 
although the formation channel via CH seems to be slightly faster
for temperatures below $\sim 7500$~K. 
The other reactions, including the direct radiative association (\#3707), are totally 
negligible compared to the CH and OH channels. 

Not only the reaction rates but rather the product 
of the rates and the number densities of the involved species is crucial for the 
change of the CO number density (see Sect.~\ref{sec:problem}). 
The average abundance of CH has a  maximum of $\sim 4.9$ in the photosphere and 
decreases above to only $\sim 3.9$, while OH is on average one to two magnitudes 
more abundant for all atmospheric heights, ranging from 
[OH]$\sim 5.0$ at the bottom of the photosphere to [OH]$\sim 7.6$ in the low 
chromosphere.  
Hence, the formation channel via OH is much more important than the CH channel 
due to the higher number density of the required reactant. 

In order to investigate this matter in more detail additional runs\footnote{
Simulation snapshots for the different cases are provided as online material.}
with different networks have been carried out. 
In all cases we started from the default network presented in 
Table~\ref{tab.reaction} and Fig.~\ref{fig.chemnetwork}. 
%
In case A all reactions involving CH are eliminated. 
The relative distribution does not change significantly. 
The absolute values are on average only reduced by $\sim9$~\% 
compared to the results with the full network. The same 
is true for the peak of the horizontally and temporally averaged CO number 
density (see Table~\ref{tab.compcases}). 
%
For case B also the reactions \#3707, \#4076, \#4097, \#7001, and \#7002 
are removed. In the remaining network carbon monoxide can only be formed 
or destroyed via OH, i.e. reactions \#14 and \#67.
The resulting CO number density closely matches the one of the previous case 
outside small regions with extreme temperatures. 
%
In contrast, the exclusion of OH (case C) produces on average by a factor of $\sim 3$ 
more CO at most positions. In chromospheric shock waves, however, $n_\mathrm{CO}$ 
is not as strongly reduced as in the case with the full network.  
Carbon monoxide is nevertheless concentrated mostly in the 
cool areas of reversed granulation in the middle photosphere. 
%
Excluding both OH and CH (case~D) results in a generally smaller CO density, where the 
largest differences are found in shock waves. 
Again, like in case~C, CO is not as strongly reduced as for the full network. 
Note that the reduced network is separated into two isolated 
parts, namely 
\mbox{H$\leftrightarrow$H$_2$} and 
\mbox{C$\leftrightarrow$CO$\leftrightarrow$O}.
%
The simulation run without the four reactions involving the representative 
metal (case E) leads basically to the same results
as the standard case. 
For temperatures below $\sim 6000$~K the difference in $n_\mathrm{CO}$ between the 
simulations with and without metal is less than one percent and 
it increases only slightly to a maximum of $\sim 2$ percent 
for higher temperatures.
We thus conclude that the adopted catalytic reactions and also the 
abundance of the representative metal are of minor importance here. 
%
Finally, the network in case~F comprises only the two reactions for radiative 
association (\# 3707) and collisional dissociation (\# 7002).  
The same pair was used in W04 \citep[see also][]{ayres96}  
but here we use the reaction coefficients chosen in this work (see Sect.~\ref{sec:network}). 
The absolute CO number density is on average roughly one order of magnitude lower 
compared to the standard case. 
The relative distribution stays quite 
similar, except for the upper layers. There the simple reaction pair produces 
a smaller relative CO concentration that, however, is not as strongly reduced 
in shock waves as it is in the case of the full network. 
This is in line with the calculations in W04 for which the 
fraction of all carbon atoms bound in carbon monoxide was only $\sim 10$~\%.
This compares to an average fraction of 80~\% derived with the full network.

\begin{table}[b]
\caption[h]{Comparison of simulations with different chemical reaction networks 
(case A-F, see text) and the full network (REF). Given are the number of remaining reactions $n_\mathrm{reac}$, absolute value ($\max(\langle n_\mathrm{CO} \rangle)$) 
and height ($z_\mathrm{max}$) of the 
maximum  of the averaged CO number density, the ratio of the peak number density and the 
maximum value of the standard case ($\delta n_\mathrm{CO} =  \max(\langle n_\mathrm{CO} \rangle)/\max(\langle n_\mathrm{CO} \rangle_\mathrm{ref})$, 
and the gas temperature $T_\mathrm{mc}$ that correlates best with CO number density. 
} 
\label{tab.compcases}
\centering
\begin{tabular}[h]{l rrrrr}
\hline
case &$n_\mathrm{reac}$&$\max(\langle n_\mathrm{CO} \rangle)$ & $z_\mathrm{max}$ & $\delta n_\mathrm{CO}$&$T_\mathrm{mc}$\\
 &&[$10^{12}\,\mathrm{cm}^{-3}$] & [km]&  &[K]\\
\hline\\[-2mm]
REF (all)           &27& 3.46& 171&   1.00& 4400\\
  A (no CH)         &19& 3.15& 185&   0.91& 4400\\
  B (no CH $+ 5$)    &14& 3.15& 185&   0.91& 4400\\
  C (no OH)         &16&10.88& 101&   3.15& 5200\\
  D (no CH, no OH)  &10& 2.37& 129&   0.69& 4700\\
  E (no M)          &23& 3.45& 171&   0.99& 4400\\
  F (\#3707,\#7002) & 2& 0.27& 157&   0.08& 4600\\
\hline
\end{tabular}
\end{table}

The ratio of the peak of the average CO number density 
of the different cases and the default network in Table~\ref{tab.compcases} 
can be used to summarise the aforementioned results. 
Catalytic reactions are negligible (case~A). 
The CH channel is only of secondary importance but should not be omitted (case~A). 
In contrast \#3707, \#4076, \#4097, \#7001, and \#7002, including 
radiative association and collisional dissociation (case~F), can be 
neglected (see case~C compared to case~B). 
Most important is the channel via OH (case~B). Excluding it from the network 
(case~C) produces fundamentally different results. 

\subsection{CO spectrum synthesis}
\label{sec.spectrum}

A sample of 50 snapshots with constant cadence from the time-dependent 2D 
simulation sequence was used as input for the spectrum synthesis code RH by 
Uitenbroek (see Sect.~\ref{sec:rhcode}). 
For all snapshots a spectrum for the same wavelength range near $4.7\ \mu\mathrm{m}$ 
was calculated for disk-centre ($\mu = 1.0$).
In Fig.~\ref{fig.spectrum} the horizontal and temporal average of these spectra 
is shown in comparison with data from the third version of the Atmospheric Trace Molecule Spectroscopy 
(ATMOS3) atlas \citep[][ \texttt{http://remus.jpl.nasa.gov/atmos}]{farmer94, farmer89}.
The ATMOS3 data were scaled to match the continuum intensity of the averaged synthetic spectrum. 

Since the CO number density of the time-dependent and the ICE approach closely match 
in the line-forming layers, also the resulting spectra agree almost perfectly. 
But both exhibit much deeper line cores than seen in the ATMOS3 data, similar 
to  the results by \citet[][ see his Fig.~11]{uitenbroek00a}. 
Owing to the dynamics, which tend to be too strong in 2D simulations, and 
the small model extent, the CO number density, the resulting depth of the absorption lines, and also 
the continuum intensity vary significantly. 
Although there are times and positions that can produce a better fit 
than the average shown here, the chosen number of 50 snapshots might still
be too small to make up a representative sample. 
This is also implied by the fact that the average continuum intensity level 
only corresponds to a brightness 
temperature of 5400~K. 
The sample is obviously not perfectly representative but rather is too cold. 
This would produce  too much CO and consequently too deep line cores. 
Lines with low excitation potential seem to deviate more from 
the ATLAS data than lines with higher excitation potential.
Given that low 
excitation lines are formed further up in the atmosphere, this trend could point 
out too much CO in the higher layers. 
On the other hand, the spectra with ICE assumption show exactly the same behaviour, 
in contrast to \citet{uitenbroek00a}. 

\section{Discussion} 
\label{sec:discus} 

\paragraph{Spatial distribution.}
As stated in Sect.~\ref{sec:spatdist} we find the maximum of the average CO height-distribution 
to be \mbox{$2.8\cdot10^{12}\,\mathrm{cm}^{-3}$} at $z = 197$~km.
Despite differences in the assumed rate coefficients, this 
is in line with the 
work by \citet{asensio03} who find, judging from their 
Fig.~2, a peak value of \mbox{$\sim 3\cdot10^{12}\,\mathrm{cm}^{-3}$} at $z \approx 100$~km. 
\citet{uitenbroek00a} does not provide a horizontal average but the spatial 
CO distribution in a two-dimensional cut through the adopted model (see his Fig.~9). 
There, the highest values are between \mbox{$\sim 3\,10^{12}\,\mathrm{cm}^{-3}$} and 
\mbox{$\sim 10^{13}\,\mathrm{cm}^{-3}$} in the low and middle photosphere. Next to these high 
values, which are mostly found above granule interiors, also smaller CO 
concentrations are present above intergranular lanes. The horizontal average 
of the distribution is thus in line with the results of this 
work, concerning the maximum values as well as its spatial distribution.  
Furthermore, the findings also agree with \citet{ayres81}, who states  
$\tau_{500} \le 10^{-2}$ as upper limit for the peak of the CO distribution. This 
optical depth corresponds to a geometrical height of $z = 365$~km  
in our model and is indeed higher than the peak found here.  
Also the fact that CO is mostly located in the cool regions of the reversed 
granulation pattern matches the observed pattern. 
\citet{uitenbroek00a} reported on spectroheliograms  in the cores of strong CO lines
near $\lambda = 4.7\,\mu\mathrm{m}$ 
that exhibit dark areas 
as small as 1\farcs, lasting for several minutes, surrounded by bright network-like 
rims \citep[see also][]{uitenbroek94}. 
The same is not only seen in the two-dimensional model presented here but even  
more clearly in  first results from 3D simulations that are currently in progress. 

The finding that the bulk of CO is truly located in the photosphere also fulfills 
other indirect constraints like the presence of  5~minute oscillations. 
Such prominent intensity oscillations were already discovered by \citet{noyes72b} 
in the core of the \mbox{3-2\ R14} CO line in the 
fundamental vibration-rotation bands ($\Delta V = 1$) 
near a wavelength of \mbox{$\lambda = 4.7\,\mu\mathrm{m}$}
\citep[see also][]{ayres90}.  
This line should be formed just in the high photosphere \citep{noyes72a}, where 
5~min oscillations can clearly be detected.

On the other hand, the differences between the synthetic spectra and the observed ATMOS3 data 
might indicate that the model still exhibits too much CO  -- at least in the 
upper layers. 
One possible cause might be a too simple treatment of the radiative transfer in the 
chromosphere which is still treated frequency-independently (grey) and under 
the assumption of local thermodynamic equilibrium (LTE). 
Already a small error in the resulting gas temperature can 
generate significant changes in CO number density. 
Another and maybe more severe source of too high CO number densities might come from incorrect
or uncertain rate coefficients. 
However, in order to answer these questions  
a larger and thus more significant set of synthetic spectra is needed.

\paragraph{Necessary modelling.}
An important result is that CO is mostly concentrated within the cool regions of 
the reversed granulation pattern. Hence, particular care should be taken for 
modelling the corresponding mid-photospheric layers. Although the basic features 
are matched, the atmospheric structure in a small two-dimensional model 
suffers from the vast dynamics which are  usually exaggerated in 2D compared to 3D. 
Consequently, we strongly recommend to proceed to three-dimensional models like,  
e.g., the one described in W04 which has been proven to produce a very realistic 
reversed granulation pattern \citep{leenaarts05}.
A 3D model provides much better statistics which are crucial for the synthesis of meaningful 
spectra (see Sect.~\ref{sec.spectrum}). 
We expect many results to remain qualitatively the same, e.g., the dependence on 
temperature and gas density (Sect.~\ref{sec:rhot}), 
the relative contribution of individual reactions (Sect.~\ref{sec:reacimport}), and 
the differences between ICE and time-dependent approach. 
In contrast the average stratification and fluctuations of temperature and CO 
number density may be a bit different in 3D compared to 2D but still the qualitative picture
should be the same. 

Another point concerns the radiative transfer which is treated grey, i.e., 
frequency-independent, in the simulations presented here. 
For comparison we calculate a short sequence with  
a non-grey multi-group scheme \citep[cf.][]{ludwig92, wedemeyer03}. 
The differences in gas temperature and in CO number 
density are still small in the middle photosphere where the 
bulk of CO is found. In contrast, there are significant differences in the 
chromosphere. The temperature fluctuations are much smaller there in the non-grey 
case. As a result the CO number densities of the grey and the non-grey simulation 
deviate most in the hottest parts of 
chromospheric shock waves. 
While the multi-group approach produces a more realistic thermal structure of 
the photosphere, the situation is less clear in case of the chromosphere. 
There, neither the grey nor the multi-group LTE approach is appropriate. 
Rather a frequency-dependent non-LTE radiative transfer is required
which is unfortunately not available yet. 
Nevertheless, the grey approach is preferable since it produces results (e.g., shock peak 
temperatures) that surprisingly agree better with other works \citep[e.g.][]{carlsson95, skartlien98}
than those obtained with the multi-group scheme (see W04 for a detailed discussion). 
But still the modelling of chromospheric radiative transfer remains a major weakness and  
thus needs to be improved in future. 

For the modelling of the number densities of the chemical species 
the ICE assumption seems to be valid in the photosphere where chemical timescales are mostly much 
shorter than the dynamical one so that advection and non-equilibrium effects 
can be neglected there -- a result already mentioned by \citet{asensio03}. 
This is also implied by the agreement of the CO distribution with the results of  \citet{uitenbroek00a}.
Hence, no time-dependent calculations seem to be necessary for the {\em photosphere}  
which after all contains by far the largest absolute amount of carbon monoxide. 
In contrast, the ICE assumption fails close to shock waves in the {\em chromosphere} 
where a time-dependent treatment is mandatory. 

The remaining discrepancies in CO equilibrium densities between the 
calculations based on the reaction network (CE) and the ICE~approach (UICE) 
can be attributed to differences in  
chemical input data
that are equilibrium constants for UICE but rate coefficients in the reaction network. 
In particular the limited temperature range for which many rates are defined  
might explain why the largest deviations are found at the hot crests of the chromospheric 
shock waves. There gas temperatures often exceed the stated range for many reactions. 

\paragraph{Chemical reactions.}
A thorough analysis of chemical reaction coefficients and additional 
simulations with altered chemical networks leads us to the conclusion 
that OH is the most important agent for forming and dissociating carbon monoxide 
whereas the CH branch contributes much less -- a result already suggested by \citet{ayres96}.
In contrast, the mere reaction pair composed of radiative association and collisional 
dissociation cannot reproduce the observed CO distribution.
Catalytic reactions involving a representative metal are negligible even if it is 
assumed to be as abundant as helium.

Unlike \citet{asensio03} we did not account for nitrogen chemistry in our
reaction network. 
In order to check the possible consequences of the exclusion of nitrogen, 
we directly compared ICE number densities 
calculated with the RH code for the exemplary time step (see Sect.~\ref{sec:results}) 
for two different sets of molecules: 
First the complete default set (see Sect.~\ref{sec:rhcode}), second a set that  
only included those 
molecules that are also part of the reaction network 
used for our time-dependent simulations. 
Hence, both sets differ by the following molecules: 
H$_2$$^+$, C$_2$, N$_2$, O$_2$, CN, NH, NO, and H$_2$O.
This includes all molecules used for nitrogen chemistry by Asensio Ramos et al..
The differences in CO number density between the two cases are neglible 
in the atmosphere, in particular at the heights where the bulk of CO is present. 
In contrast, the reduced set resulted in a CO number density lower by up to five 
orders of magnitude at the hot centres of chromospheric shock waves but there only little CO 
is found anyway. 
Moreover, the average CO height distribution of our time-dependent simulation is very
similar to that of Asensio Ramos and co-workers. 
We thus conclude that nitrogen chemistry may be negligible for the formation of CO 
in  the solar atmosphere -- at least outside the problematic hot centres of chromospheric 
shock waves.

\paragraph{Chemical timescales.}
The rates and the corresponding timescales span a range of 
several orders of magnitude, making it difficult to define a meaningful average. 
The results presented here should thus serve as a first-order 
proxy only. 

\section{Conclusions} 
\label{sec:conclusion} 

The presented 2D radiation chemo-hydrodynamic simulations
produce a spatial distribution of carbon monoxide that agrees well with earlier 
theoretical work and also with observations. 
The synthetic spectra match reasonably well if one considers the large 
fluctuations that are unavoidable in the case of such a small-size  
two-dimensional model. 
More accurate results can be expected from a 3D simulation which is
currently in progress. 
The presented model proved that CO is mostly located in the 
cool regions of the reversed granulation pattern at mid-photospheric heights.
Some effort was spent to make the applied chemical reaction network as realistic as possible 
on the basis of the available  chemical input data whose quality is quite poor 
in some cases. Hydroxide (OH) was found to be the most important ingredient for 
the formation and destruction of carbon monoxide.

The presented comparison between the time-dependent
simulation and the corresponding instantaneous chemical 
equilibrium (ICE) case shows that significant deviations occur 
only near hot shock waves in the solar chromosphere. 
Hence, ICE is a valid assumption for the photosphere
that allows to treat CO in a simple way in simulations that do not extend beyond 
the photosphere. 
In contrast, ICE fails in and close to propagating high-temperature regions 
at chromospheric densities. 
Rather a time-dependent treatment, as presented in this work, is mandatory for 
realistic simulations of the solar chromosphere. 

Last but not least we take the agreement of the results with other works as 
demonstration of the validity of the approximations and algorithms of the upgraded code 
\textsf{\mbox{CO$^5$BOLD}}
in the case of CO in the solar atmosphere. 
It now can be used for a large range of applications for other stellar types and 
also other chemical reaction networks. 

\begin{acknowledgements} 
T.~Ayres, A.~Konnov, and T.~J.~Millar gave helpful advise concerning chemical 
reaction rates. 
In addition we thank T.~Ayres for valuable comments and H.~Uitenbroek for 
providing his RH code.  
SW acknowledges support by the 
{\em Deutsche Forschungs\-gemein\-schaft (DFG)}, 
project Ste~615/5. 

\end{acknowledgements} 


\end{document}